\documentclass[onecolumn,amsmath,amssymb]{revtex4}

\def\lsim{\raise0.3ex\hbox{$\;<$\kern-0.75em\raise-1.1ex\hbox{$\sim\;$}}}
\def\gsim{\raise0.3ex\hbox{$\;>$\kern-0.75em\raise-1.1ex\hbox{$\sim\;$}}}

\usepackage{graphics}

\newcommand{\be}{\begin{eqnarray}}
\newcommand{\ee}{\end{eqnarray}}

\def\bea{\begin{eqnarray}}
\def\eea{\end{eqnarray}}

\usepackage{epsfig}
\usepackage{array}
\usepackage{booktabs}

\begin{document}
\title{$B-L$ heavy neutrinos and neutral gauge boson $Z'$ at the LHC}
\author{A. A. Abdelalim$^{1,2}$, A. Hammad$^{1}$ and  S. Khalil$^{1,3}$}
\vspace*{0.2cm}
\affiliation{$^1$ Center for Fundamental Physics, Zewail City for Science and Technology, 6 October 
$^2$Department of Physics, Faculty of Science,  Helwan University, Cairo, Egypt. \\
$^3$Department of Mathematics, Faculty of Science,  Ain Shams University, Cairo, Egypt. }
\date{\today}

\begin{abstract}

We explore possible signatures for heavy neutrinos and neutral gauge boson, $Z'$, in TeV scale $B-L$ 
extension of the Standard Model (BLSM) with inverse seesaw  mechanisms at the Large Hadron Collider.  
We show that due to new decay channels of $Z'$ into heavy/inert neutrinos,  the LHC stringent bounds 
imposed on $Z'$ mass can be significantly relaxed. We analyze  the  pair production of heavy neutrinos decaying to 
four leptons plus two neutrinos, four jets plus two leptons, or three leptons plus two jets and 
one neutrino.  We show that the $4 l + 2 \nu$ is the most promising decay channel for probing both $Z'$ and heavy neutrinos at the LHC.  

\end{abstract}
\maketitle


The solid evidence for neutrino oscillation, pointing towards non-vanishing neutrino masses, is one 
of the firm hints for physics beyond the Standard Model (SM).
Neutrinos are strictly massless in the SM due to two main reasons: $(i)$ the absence of right-handed 
neutrinos, $(ii)$ the SM has an exact global Baryon minus Lepton (B-L) number conservation. 
The minimal extension of the SM, based on the gauge group $SU(3)_C \times SU(2)_L 
\times U(1)_Y \times U(1)_{B-L}$, can account for the light neutrino masses through either Type-I 
seesaw or Inverse seesaw (IS) mechanism \cite{Khalil:2006yi,Khalil:2010iu}. In type-I seesaw mechanism 
right-handed neutrinos acquire Majorana masses at the $B-L$ symmetry breaking scale, which can be 
related to the supersymmetry breaking scale, {\it i.e.}, ${\cal O}(1)$ TeV \cite{Khalil:2007dr}. While in IS, 
these Majorana masses are not allowed by $B-L$ gauge symmetry and another pair of SM gauge singlet 
fermions with tiny masses $(~ {\cal O}(1)$ keV) must be introduced. One of these two singlets fermions 
couples to right handed neutrino and is involved in generating the light neutrino masses. The other 
singlet (is usually called inert neutrino) is completely decoupled and interacts only
through the $B-L$ gauge boson, therefore it may account for Warm Dark Matter\cite{Basso,El-Zant:2013nta} .
In both scenarios, this model predicts several testable signals at the LHC through the new predicted 
particles: $Z'$ (neutral gauge boson associated with the $U(1)_{B-L}$), extra Higgs (an additional singlet 
state introduced to break the gauge group $U(1)_{B-L}$ spontaneously), and three (type-I) or six (IS) 
heavy neutrinos, $\nu_h$, that are required to cancel the associated anomaly and are necessary for the 
consistency of the model.

In this paper we aim to provide a comprehensive analysis for the LHC potential discovery of $Z^\prime$ 
and $\nu_h$'s predicted in BLSM with IS neutrino mechanisms.  We show that the possibility $Z'$ decay 
into a pair of heavy/inert neutrinos is salient feature of this class of model and provides a very important 
signature for probing both $Z'$ and $\nu_h$ of BLSM at the LHC.  Due to the presence of these channels, one finds that
the branching ratio of $Z' \to l^+ l^-$ is suppressed with respect to the one in other models of $Z'$, like 
what is called Sequential SM (SSM), which is usually considered as bench mark in experimental searches 
for $Z'$ gauge boson \cite{AtlasResult, cmsResult}. Therefore, the recent LHC stringent bounds imposed 
on $Z'$ mass can be relaxed in BLSM.  We also investigate the LHC discovery potential for the heavy neutrinos in BLSM through its decay into leptonic, hadronic and semi-leptonic decay channels.  
We provide a phenomenological study for the decay channels with four leptons plus two $SM-$like neutrinos, four jets plus two leptons, and three leptons plus two jets and one $SM-$like neutrino.  We show that the decay of $Z'$ into two heavy neutrinos that decay to four hard leptons and large missing energy due to the associated neutrinos is very clean with negligible SM background. It is important to mention that our analysis is a completion of the previous work on Type-I BLSM \cite{Khalil:11,Khalil:1} and the first of its kind in analyzing the phenomenological implications of inverse seesaw BLSM.  


In BLSM with IS mechanism, one assumes that the SM singlet scalar $\chi$, which spontaneously breaks 
$U(1)_{B-L}$, has $B-L$
charge $=-1$. Also, three pairs of SM singlet fermions: $S_{1,2}$ with $B-L$ 
charge $=\mp 2$, respectively, are introduced.
Therefore, the corresponding Lagrangian of leptonic sector is given by
\cite{Khalil:2010iu}
\begin{eqnarray}
{\cal L}_{B-L}&=&-\frac{1}{4} F'_{\mu\nu}F'^{\mu\nu} + i~
\bar{\ell}_L D_{\mu} \gamma^{\mu} \ell_L + i~ \bar{e}_R D_{\mu}
\gamma^{\mu} e_R \nonumber\\
&+& i~ \bar{\nu}_R D_{\mu} \gamma^{\mu} \nu_R + i~ \bar{S}_{1} D_{\mu} \gamma^{\mu} S_{1} + i~ 
\bar{S}_{2}
D_{\mu} \gamma^{\mu} S_{2} \nonumber\\
&+&(D^{\mu}\phi)^\dagger D_{\mu} \phi + (D^{\mu} \chi)^\dagger D_{\mu}\chi -V(\phi, \chi)
\nonumber\\
&-&\Big(\lambda_e \bar{\ell}_L \phi\, e_R+\lambda_{\nu}
\bar{\ell}_L \tilde{\phi} \nu_R +\lambda_{S} \bar{\nu}^c_R \chi
S_2 \Big) + h.c..~~~
\label{lagranigan}
\end{eqnarray}
After the $B-L$ and the EW symmetry breaking, through non-vanishing vacuum expectation 
values (VEVs) of $\chi$:
$|\langle\eta\rangle|= v'/\sqrt 2$ and $\phi$:
$|\langle h \rangle|= v/\sqrt 2$, one finds that the neutrino
Yukawa interaction terms lead to the following mass
terms\cite{Khalil:2010iu}:%
\be%
{\cal L}_m^{\nu} = m_D \bar{\nu}_L \nu_R + M_N \bar{\nu}^c_R S_2 + h.c.,%
\ee%
where $m_D=\frac{1}{\sqrt{2}}\lambda_\nu v$ and $ M_N =
\frac{1}{\sqrt 2}\lambda_{S} v' $. Here $v'$ is assumed to be of
order TeV and $v =246$ GeV. Moreover, one may generate very small Majorana masses for
$S_{1,2}$ fermions through possible non-renormalizable terms like
$\bar{S}^c_{1} {\eta^\dag}^{4} S_{1}/M^3$ and $\bar{S}^c_{2}
{\eta}^{4} S_{2}/M^3$. Hence, the Lagrangian of neutrino masses, in the flavor basis, is given by %
\be%
{\cal L}_m^{\nu} =\mu_s \bar{S}^c_2 S_2 +(m_D \bar{\nu}_L \nu_R + M_N \bar{\nu}^c_R S_2 +h.c.) ,%
\ee%
where $\mu_s=\frac{v'^4}{4 M^3}\lsim 10^{-6}$ GeV. Therefore, the
neutrino mass matrix can be written as ${\cal M}_{\nu}
\bar{\psi}^c \psi$
with $\psi=(\nu_L^c ,\nu_R, S_2)$ and ${\cal M}_{\nu}$ is given by %
\be {\cal M}_{\nu}=
\left(%
\begin{array}{ccc}
  0 & m_D & 0\\
  m^T_D & 0 & M_N \\
  0 & M^T_N & \mu_s\\
\end{array}%
\right). %
\label{inverse}
\ee%
Note that in order to avoid a possible large mass term $m S_1 S_2$ in the Lagrangian Eq.(\ref{lagranigan}), 
that would spoil the above inverse seesaw structure, one assumes that the SM particles, $\nu_R$, $\chi$, 
and $S_2$ are even under a $Z_2$-symmetry, while $S_1$ is an odd particle. Also other discrete 
symmetry may be used to avoid other possible non-renormalizable terms \cite{Waleed}. 

The diagonalization of the mass matrix Eq.(\ref{inverse}) leads to the following light and heavy neutrino 
masses, respectively: %
\begin{eqnarray}%
m_{\nu_l} &=& m_D M_R^{-1} \mu_s (M_R^T)^{-1} m_D^T,\label{mnul}\\
m_{\nu_H}^2 &=& m_{\nu_{H'}}^2 = M_R^2 + m_D^2. %
\end{eqnarray} %
Thus, one finds that the light neutrino masses can be of order eV, with a TeV scale $M_R$ if $\mu_s \ll 
M_R$ and
order one Yukawa coupling $\lambda_{\nu}$. Such large coupling is crucial for testing the BLSM with 
inverse seesaw and probing the heavy neutrinos at the LHC.
From Eq. (\ref{mnul}), one finds that the $9\times 9$ neutrino mass matrix
${\cal M}_\nu$ can be diagonalized by the matrix $V$, {\it i.e.}, $V^T {\cal M}_\nu V = {\cal
M}_\nu^{\rm diag}$ \cite{Waleed}, where %
\be V=
\left(%
\begin{array}{cc}
  V_{3\times 3} & V_{3\times6}\\
  V_{6\times 3} & V_{6\times6}  \\
\end{array}%
\right),%
\ee%
with $V_{3\times3}$ is given by %
\be%
V_{3\times3} \simeq \left(1-\frac{1}{2} F F^T \right)
U_{\!M\!N\!S} \, . %
\ee%
The
matrix $V_{3\times6}$ is defined as %
\be%
V_{3\times6}=\left({\bf 0}_{3\times3},F \right) V_{6\times6},\quad
~~~  F = m_D M^{-1}_R. %
\label{V36}
\ee %
Finally, $V_{6\times 6}$ is the matrix that diagonalize the
$\{\nu_R, S_2\}$ mass matrix. In order to grantee that the first three eigenvalues of
the light neutrino mass matrix ${\cal M}_\nu$ are consistent with the physical light neutrinos,
one writes the Dirac neutrino mass matrix $m_D$ as :%
\be %
m_D=U_{\!M\!N\!S}\, \sqrt{m_{\nu_l}^{\rm diag}}\, R\, \sqrt{\mu^{-1}_s}\, M_N, %
\label{mD}
\ee %
where $R$ is an arbitrary orthogonal matrix.

As shown in reference \cite{Waleed}, the mixings between light and heavy neutrinos are of order 
${\cal O}(0.01)$. Therefore, the decay
width of these heavy neutrinos into SM fermion are sufficiently large. It is worth to mention that 
the second SM-singlet fermion, $S_1$, remains light with mass given by%
\be%
m_{S_1} = \mu_s \simeq {\cal O}(1)~ {\rm keV}.%
\ee%
where $S_1$ is a kind of inert neutrinos that has no mixing with active neutrinos. It can be a good 
candidate for warm dark matter as emphasized in Ref.\cite{El-Zant:2013nta}.

\begin{figure}[t]
\epsfig{file=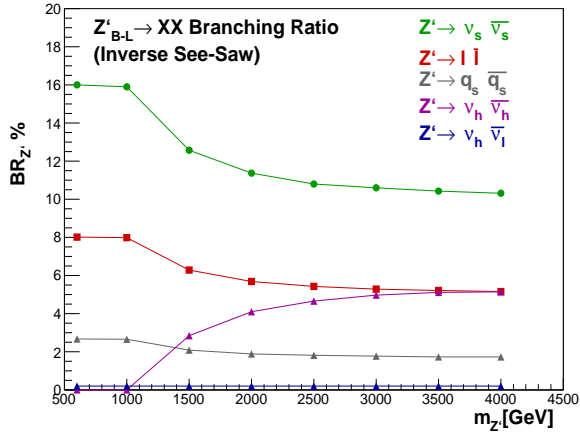,height=5.9cm,width=7.9cm,angle=0}
\caption {Branching Ratios of $Z^\prime$ decays in BLSM with IS as function of $M_{Z'}$. } 
\label{fig1}
\end{figure}

Now we study the signatures of the extra neutral gauge boson $Z^\prime$ in BLSM with IS mechanisms at LHC.  The possibility of $Z^\prime$ decay into a pair of heavy/inert neutrinos would enlarge the total decay width of $Z'$. Therefore, the $BR(Z' \to l^+ l^-)$  is suppressed with respect to the prediction of the Sequential Standard Model (SSM), which is usually considered as benchmark for the experimental search of $Z'$. Fig. \ref{fig1} shows the branching ratios of all $Z^\prime$ decays.
According to this figure, the branching ratios of $Z^\prime$ decays are given by%
\bea
&&\sum_{l} BR (Z^\prime \rightarrow l \bar{l}+v_l \bar{v_l}) \sim  20.4\% \nonumber\\
&&\sum_{q} BR(Z^\prime \rightarrow q \bar{q}) \sim 9.2\% \nonumber\\
&&\sum_{\nu_h} BR (Z^\prime \rightarrow \nu_h \bar{\nu}_h) \sim 24\% \nonumber\\
&&\sum_{\nu_s} BR (Z^\prime \rightarrow \nu_s \bar{\nu}_s) \sim 41.4\% .
\eea
where $l=e$ or $\mu$, $q = u, d, s,$ or $c$, while $\nu_h$ refers to the six heavy neutrinos, and $\nu_s$ refers to the three inert neutrinos.

It is worth noting that in our model the $Z^\prime$ cross sections that were used to derive the ATLAS and CMS current mass limit could be simply rescaled by a factor of $(g_{B-L}/g_{Z})^2 \times (1 - BR(Z^\prime\ decay\ new\ channels))$. If $g_{B-L}=g_Z$ and $BR(Z^\prime\ decay\ new\ channels)=0$, this reproduces the SSM cross-sections that were used by ATLAS and CMS. Considering the scaling of cross-sections, the current $Z^\prime$ mass limits will be lowered by a factor of $\sigma_{B-L}(Z^\prime \to ll)/\sigma_{SSM}(Z^\prime \to ll)$. This result is consistent with the conclusion of Ref.\cite{Arcadi:2013qia}.
 
 If $M_Z^\prime = 1000 GeV$ were considered,  $BR(Z' \to  l^+ l^-) \sim 14\%$ and the $\sigma \times BR = 16 fb$ when $g_{B-L} = g_Z=0.188$ and the $\sigma\times BR = 82 fb$ when $g_{B-L} = 0.5$, while in the SSM the $BR(Z' \to  l^+ l^-) \sim 7.6\%$ that gives $\sigma\times$BR $= 340 fb^{-1}$ for both electron and muon channels. In this respect, the experimental limits: $M_{Z'}  \gsim 2.5$ TeV \cite{AtlasResult} ($2.8$ TeV) \cite{cmsResult} will be lowered to $0.247$ of its value when $g_{B-L} = 0.5$. Such a lower $Z'$ mass is sticking signature of BLSM with IS. In other scenario like Type I BLSM, the $BR(Z' \to  l^+ l^-) \sim 28.6\%$ and $\sigma \times BR = 814 fb$ that leads to increasing the current mass limit at $g_{B-L} = 0.5$. Table \ref{tabXsec} gives $\sigma\times$BR($Z^\prime\to ee$) for SSM and BSLM with IS at different $g_{B-L}$. Fig. \ref{fig2} shows the invariant mass of the di-lepton from $Z^\prime$ decay to lepton pair in IS BLSM (blue) and in SSM (red) with 20 $fb^{-1}$ at $\sqrt{s}= 8$ TeV, where lepton here refer to electron only and considering $Z^\prime$ mass of 1000 GeV as a benchmark point.

\begin{table}[h]
\caption{\emph{$Z^\prime\to ee$ cross sections times branching ratios at different masses.}}
\centering
\begin{tabular}{*5l}
\hline
 $M_{Z^\prime}$ [GeV]    &  $\sigma_{SSM}$ [$fb$]  &  \multicolumn{3}{c}{ $\sigma_{B-L}$  [$fb$] with IS}                                   \\
                               &                           &  $g_{B-L}=g_Z\ \ $ & $g_{B-L}=0.5\ \ $   & $g_{B-L}=0.8$ \\                               
\hline
\hline
1000                      & 170  & 6            &  41 & 105.7 \\
\hline
1500                      & 21.7  & 0.58      &  4.5 & 13.2 \\
\hline
2000                      & 3.4  & 0.087      &  0.72 & 2.3 \\
\hline
2500                      & 0.8  & 0.015      &  0.15 & 0.58 \\
\hline
3000                      & 0.21  & 0.003      &  0.04 & 0.19 \\
\hline
\end{tabular}
\label{tabXsec}

\end{table}

\begin{widetext}
\begin{center}
\begin{figure}[h]
\epsfig{file=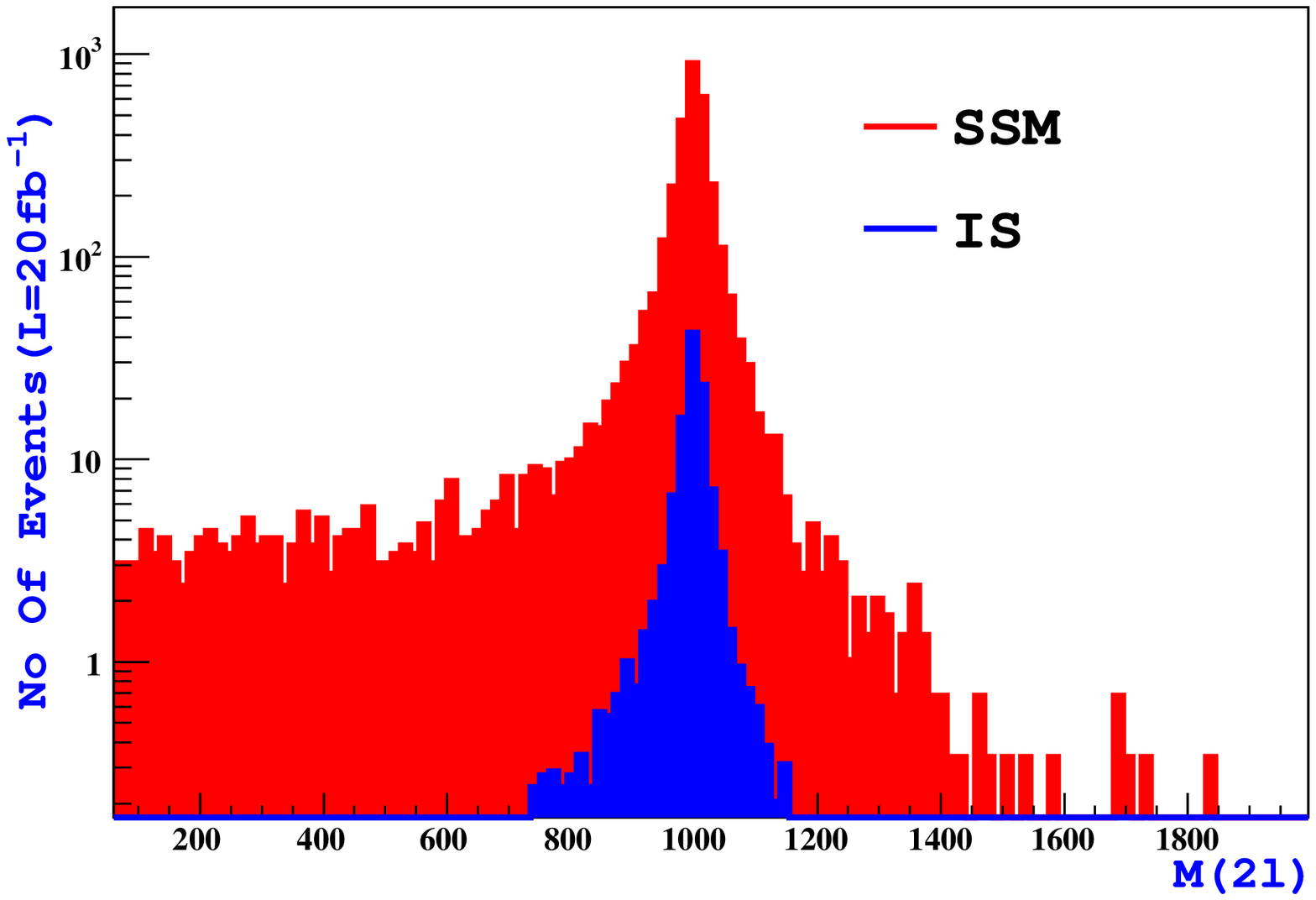,height=5.0cm,width=8.0cm,angle=0}~~~\epsfig{file=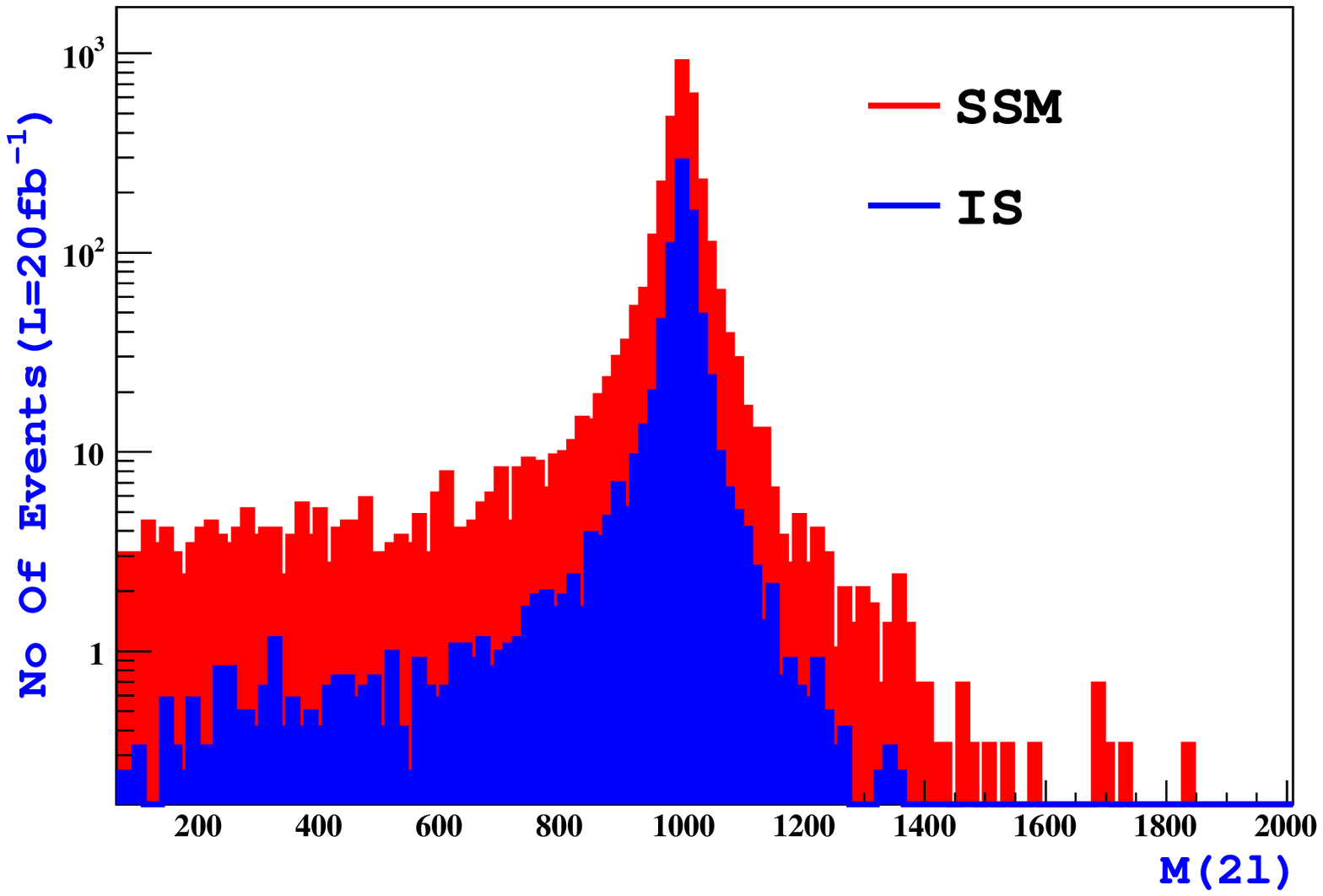,height=5.0cm,width=8.0cm,angle=0}
\caption {Invariant mass of dielectron from $Z^\prime$ decaying to electron pair in 
case of SSM (red) and in case of IS BLSM (blue) at $g_{B-L}= g_{Z}$ (left) and $g_{B-L}=0.5$ (right).}
\label{fig2}
\end{figure}
\end{center}
\end{widetext}

\begin{figure}[h]
  \centering
  \includegraphics[height=3.5cm,width=4.5 cm,scale=0.1]{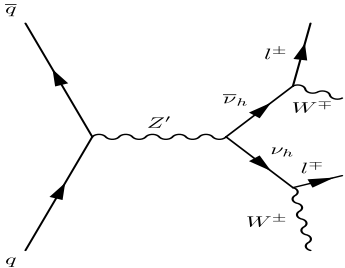}
  \caption {$ q\bar{q} \rightarrow Z^\prime \rightarrow \nu_h \bar{\nu_h} \rightarrow WW ll $ Feynman diagram}
    \label{figFeynman}
\end{figure}

The dominant production mode for the heavy neutrinos at the LHC would be through the Drell-Yan mechanism, with $Z^\prime$. The mixing between light and heavy neutrinos generates new couplings between the heavy neutrinos, the weak gauge boson $W$ and $Z$, and the associate leptons. These couplings are crucial for the decay of the heavy neutrinos. The main decay channel is through $W$ gauge boson, which may decay leptonicaly or hadronicaly, a Feynman diagram is given in Fig. \ref{figFeynman}. In case of multi-leptons final state one ends with four leptons plus missing energy ($4l+ 2\nu_l$), while in case of multi-hadronic final state states one ends with four jets plus two leptons ($4J+2l$). In addition, it is also possible to have mixed final with ($3l + 2j+ \nu_l$). If two flavors of the heavy neutrinos are assumed to be degenerate in mass, one gets the same final states for the produced heavy neutrino pair with similar event rates. This will double the number of final state events but on the other hand, it makes it difficult to distinguish between final state leptons. Therefore through out the current study, we considered the non-degenerate heavy neutrino masses considering the interference between every two different flavor.

In our analysis we have used SARAH\cite{SARAH} and SPheno\cite{Spheno} to build the model. The matrix-element calculation and events generation are derived by MadGraph\cite{madgraph}. Finally we used Pythia\cite{Pythia} to simulate the initial and final state radiation, fragmantation and hadronization effects. We considered the following bench mark: $M_{Z^\prime} = 1000\ GeV$, $M_{\nu_4}=M_{\nu_5}= 287\ GeV$, $M_{\nu_6}=M_{\nu_7}= 435\ GeV$, and $M_{\nu_8}=M_{\nu_9}= 652\ GeV$.  In addition, the following cuts are assumed: A lower transvers-momentum, $p_T$, cut of $20\ GeV$ ($10\ GeV$) was set on final state jets (electrons) and a higher pesudo-rapidity, $\eta$, cut of 4 (2) was set on jets (electrons), finally the separation between two jets (electrons), $R_{jj}\ (R_{ll})$ was set to be 0.4 (0.2).     


{\it i) 4 l + 2 $\nu$ Final State}: the main advantage of this channel is that it is almost background free. The main SM background comes from the three gauge boson $WWZ$ production with $\sigma(WWZ)\sim 200\ fb$ at 14 TeV\cite{Barger:1988}. In Fig. \ref{fig4} we show the generator level invariant mass of 4-leptons plus 2 light neutrinos from $Z^\prime$ with $WWZ$ background and also the invariant mass of 2-leptons plus light neutrino from heavy neutrino decay. In the right plot, it is clear that the heaviest two neutrinos ($\nu_8$ and $\nu_9$) are decayed off-shell when the $M_{Z^\prime}= 1000\ GeV$. These figures indicate that the decay channel {\it 4 l + 2 $\nu$} is the quite clean channel and quite promising for probing both $Z^\prime$ and $\nu_h$ using only few cuts to extract signals from background. The number of event left after set of cuts mentioned above are 270 signal events and 10 background events. 

In Fig. \ref{lumiForDisc4l} (left panel) we display the Integrated Luminosity of the data need for 1-7 $\sigma$ Statistical Significance discovery for  $g_{B-L}=0.5$ and $M_{Z'}$=1000, 1500 and 2000 GeV. In the right panel while the left plot we plot the Integrated Luminosity of the data need for a 5 $\sigma$ discovery as a function of $M_{Z'}$ for $g_{B-L} = 0.5$ and $g_{B-L} = 0.8$. If one consider the case of $g_{B-L} = 0.188$, which corresponds to the case of SSM, he finds that the Luminosity needed for a 5 $\sigma$ discovery is of order 390 $fb^{-1}$ for $M_{Z'}= 1000$ GeV and about $8.4 \times 10^5$ for $M_{Z'}= 2500$ GeV. This value of Luminosity is incredibly high and far beyond the expected data of the LHC experiments during the foreseen Run II.  Therefore, one may conclude that the scenario of SSM is rather unfavored not only for the reason of using $Z^\prime$  decays to di-leptons channel only, but also for using SM like value for coupling. 

\begin{widetext}
\begin{center}
\begin{figure}[h]
\epsfig{file=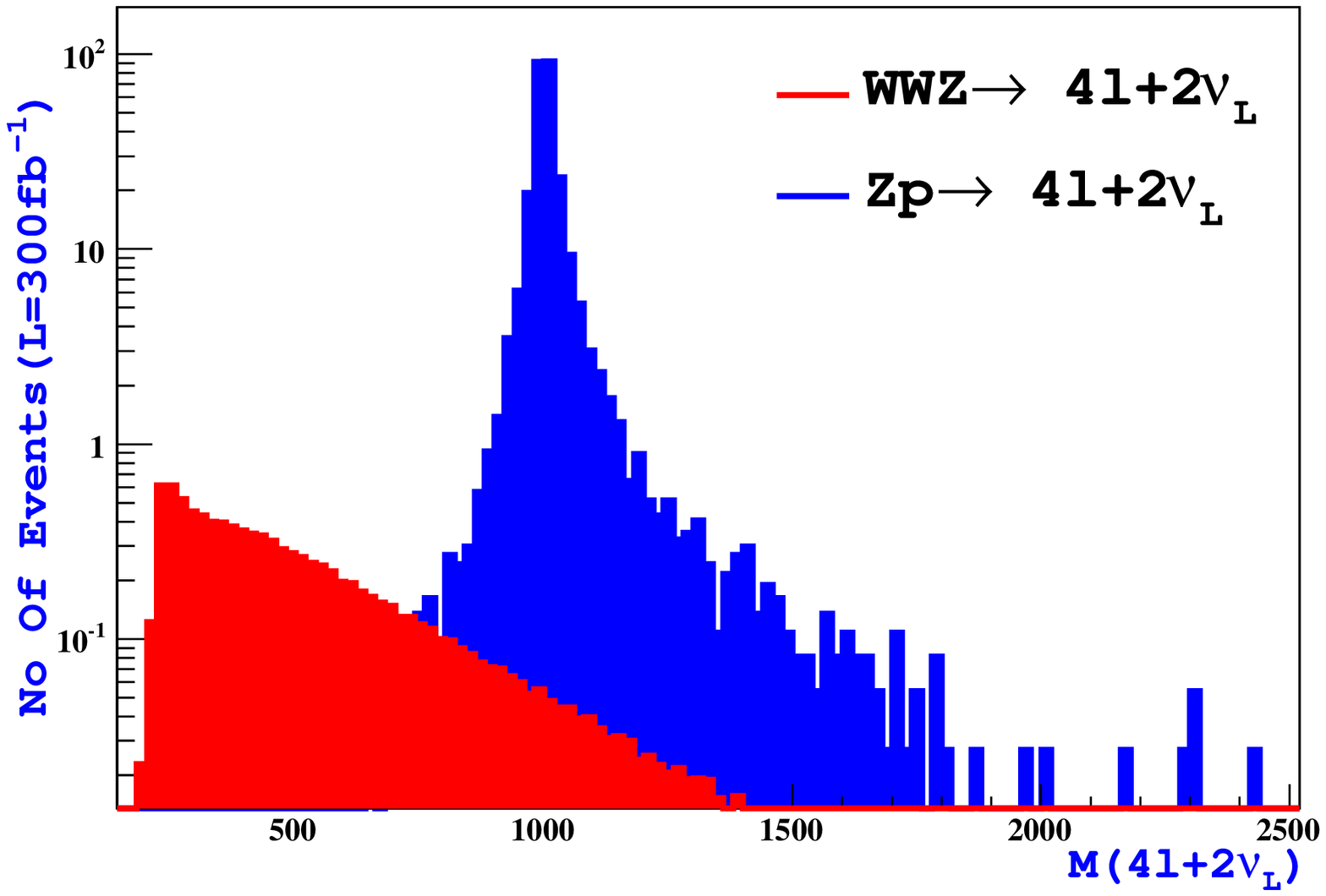, height=5.0cm,width=8.0cm,angle=0}~~~ \epsfig{file=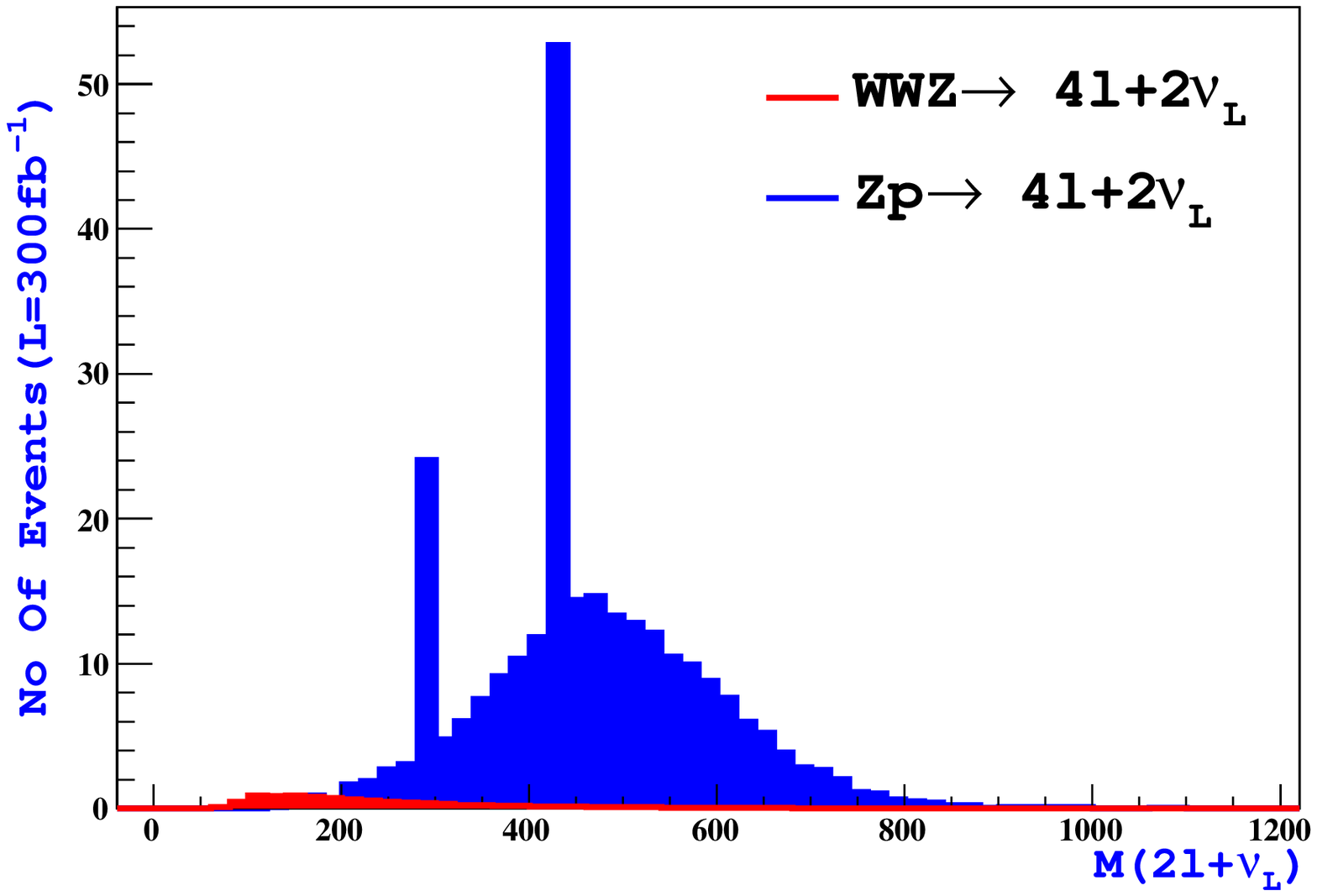,height=5.0cm,width=8.0cm,angle=0}
\caption {(Left) The invariant mass of the 4-leptons plus 2 light neutrinos distribution from $Z'$ decay. (Right)
The invariant mass of 2-leptons plus 1 light neutrino from heavy neutrino decay. Expected SM background is included as well. }
 \label{fig4}
\end{figure}
\end{center}
\end{widetext}

\begin{widetext}
\begin{center}
\begin{figure}[h]
\epsfig{file=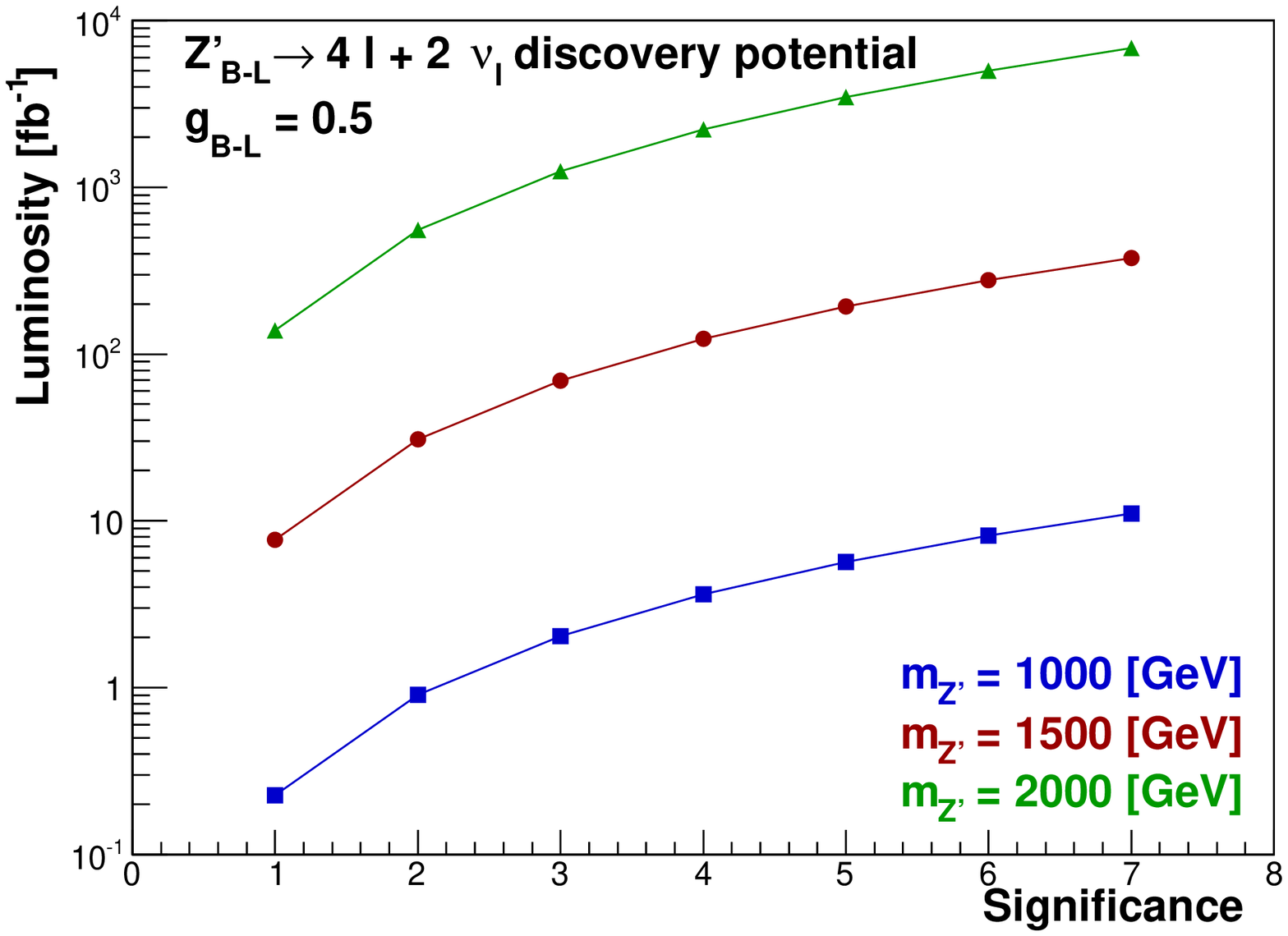, height=5.0cm,width=8.0cm,angle=0}~~~ \epsfig{file=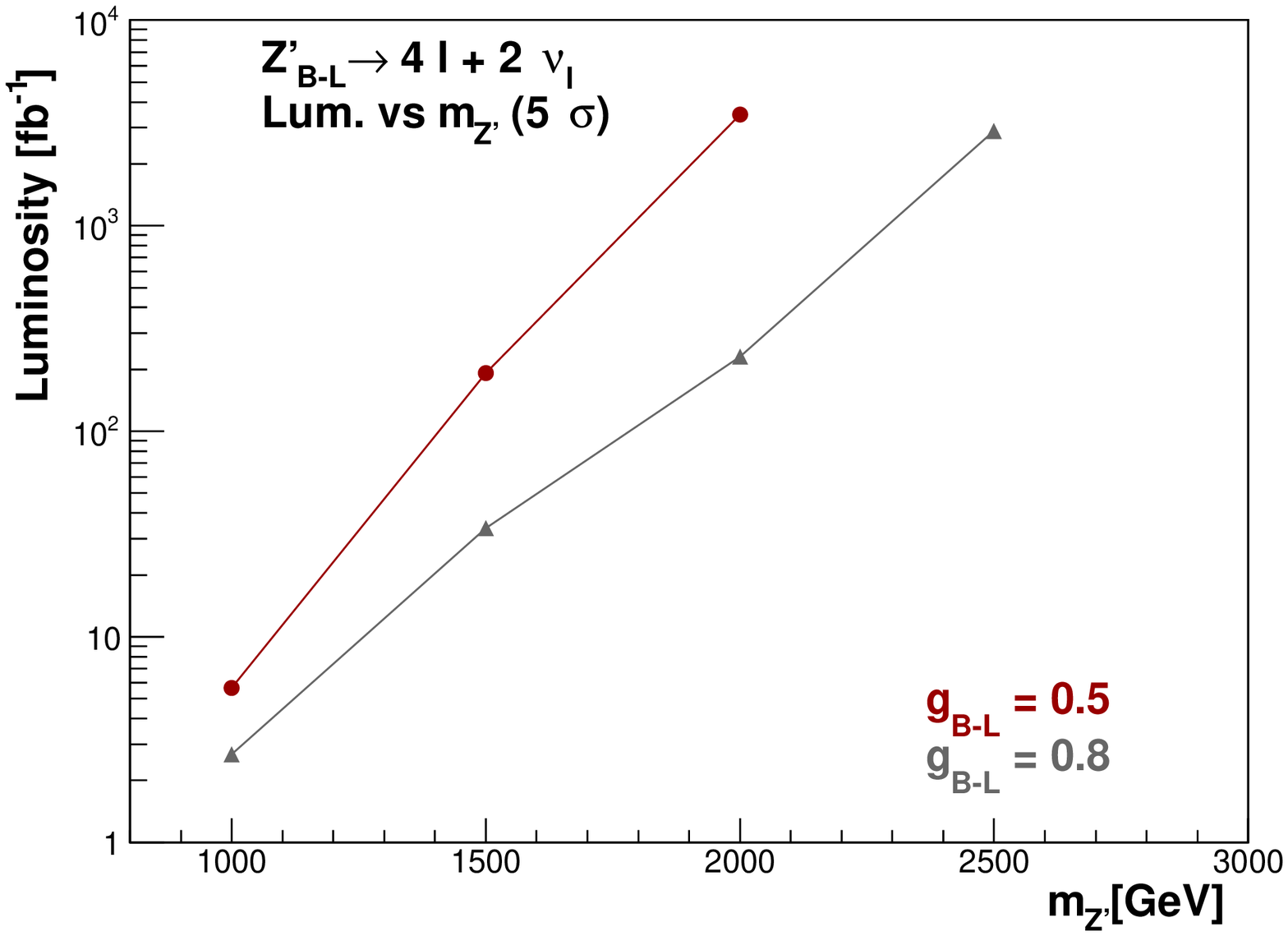,height=5.0cm,width=8.0cm,angle=0}
\caption {For the {\it4 l + 2 $\nu$} final state, the Integrated Luminosity of the data need for (Left) 1-7 $\sigma$ Statistical Significance discovery at  $g_{B-L}=0.5$ for different $M_{Z'}$, (Right) a 5 $\sigma$ discovery as a function of $Mass_{Z'}$ at $g_{B-L} = 0.188$, $g_{B-L} = 0.5$ and $g_{B-L} = 0.8$. }
 \label{lumiForDisc4l}
\end{figure}
\end{center}
\end{widetext}


{\it ii) 4 j + 2 l Final State}: here both $W$'s decay hadronicaly and because of the higher branching ratio of $ W \rightarrow j j^\prime \sim 60 \%$, we expect higher number of events than the previous channel ({\it 4 l + 2 $\nu$}). The irreducible SM background is due to $Z Z + jj$, where one of the $Z$ decays to two leptons while the other $Z$ decays two quark anti-quark ($Z \rightarrow l \bar{l}, Z \rightarrow j \bar{j}$). The contribution due to $Z+jets$ could be neglected. In addition, there are two reducible background coming from $t \bar{t}$ and $W W$. In figure \ref{fig5} we present invariant mass of $4 j + 2 l$ (left) and also that of $2 j + l$ from signal and background, after applying an additional $p_T$ cut of 150 GeV on the two $p_T$-leading leptons. As in the $4 l + 2 \nu$ channel, the heaviest two neutrinos ($\nu_8$ and $\nu_9$) are decayed off-shell when the $M_{Z^\prime}= 1000\ GeV$. The number of event left after cuts for signal and backgrounds are listed in table \ref{table:4jets}. 
The Integrated Luminosity of the data need for 1-7 $\sigma$ Statistical Significance discovery is shown in Fig. \ref{lumiForDisc4j} (left panel) at  $g_{B-L}=0.5$ and $M_{Z'}$= 1000, 1500, 2000 and 2500 GeV. The presence of four jets in the final state of the signal, makes SM backgrounds highly contributed. With a cut $P_T > $ 150 GeV, the signal is fully enhanced over SM backgrounds which make the 5 $\sigma$ discovery predictable in the near future at LHC as shown in Fig. \ref{lumiForDisc4j} (right panel).

\begin{widetext}
\begin{center} 
\begin{figure}[h]
\epsfig{file=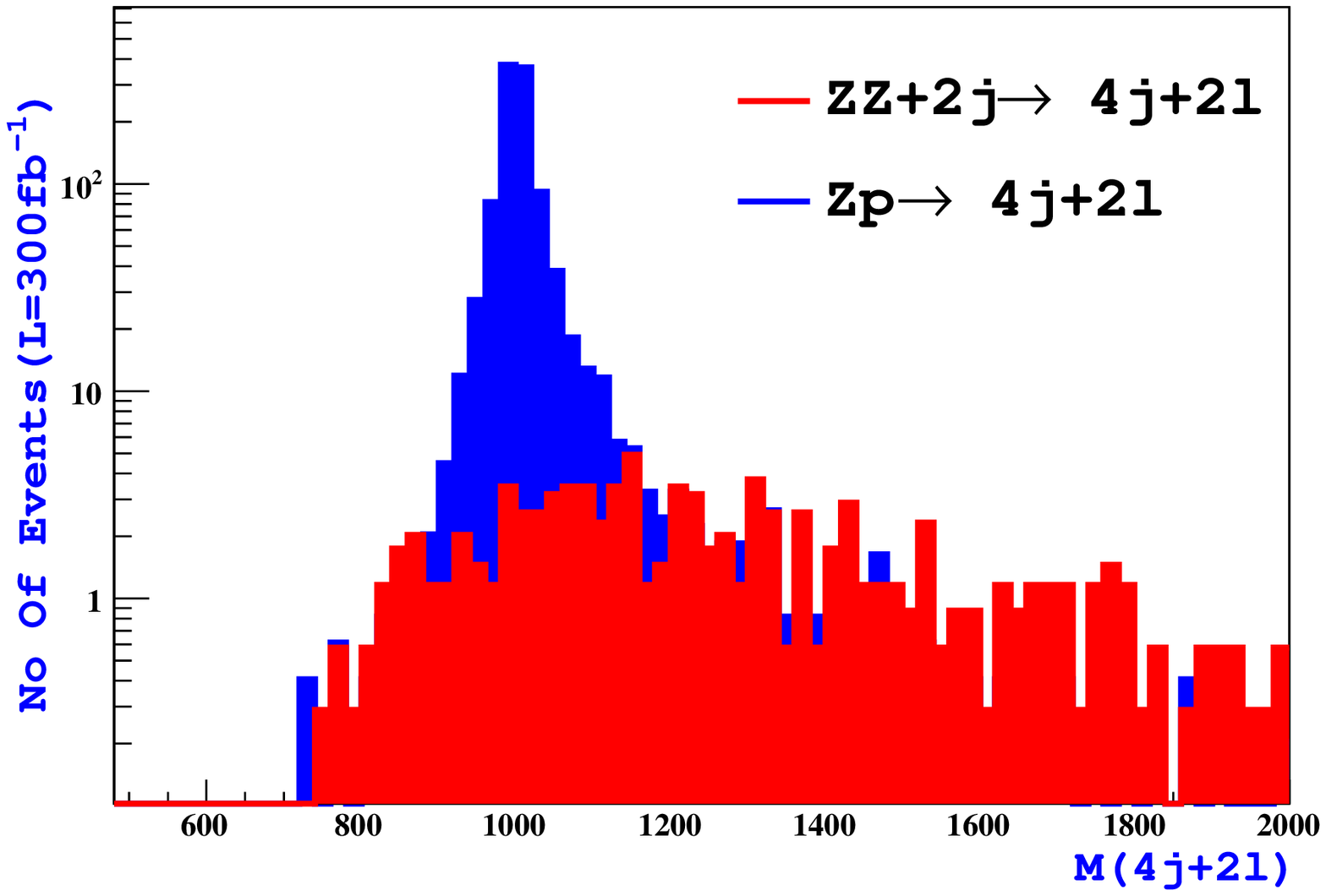,height=5.0cm,width=8.0cm,angle=0}~~
\epsfig{file=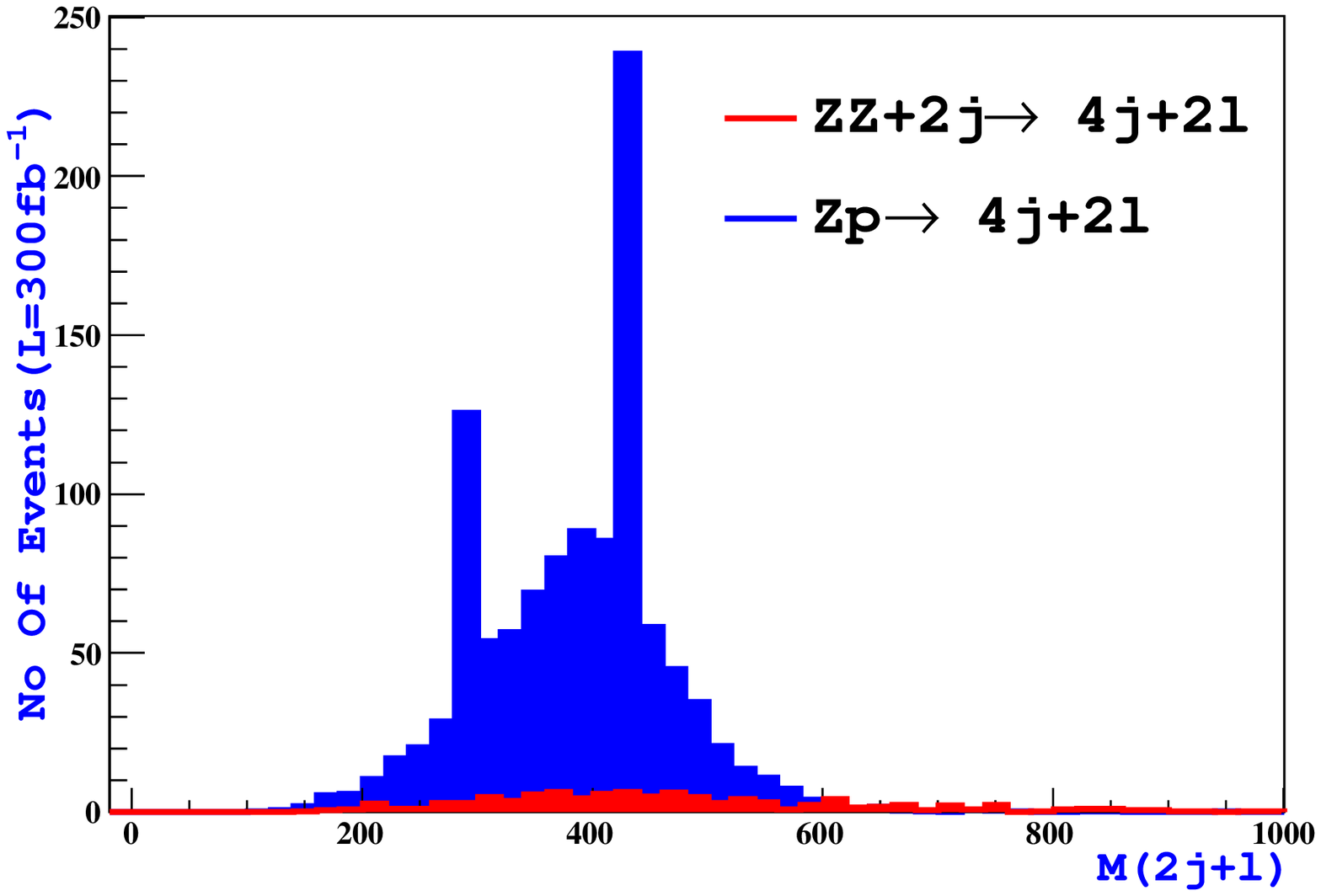,height=5.0cm,width=8.0cm,angle=0}
  \caption {(Left) The invariant mass of the 4 jets plus 2 leptons distribution from for signal as well as expected SM background. (Right)
The invariant mass of 2 jets plus 1lepton from heavy neutrino decay.}
  \label{fig5}
\end{figure}
\end{center}
\end{widetext}
\begin{table}[ht]
\caption{Number of events after initial set of cuts, in case $Z^\prime \to 4j+2l$ }
\centering
\begin{tabular}{c c c c c}
\hline\hline
Cut & Signal  & ZZjj & t$\bar{t}$ & WW  \\
\hline
Initial No of events \footnote{Weighted events, the initial generated number of events is 10 K events in signal and 100 K events for background sample} &
2042   & 28913 & 650000 & 80000 \\
$p_T > 150 GeV$ & 1088 & 102 & 0 & 0  \\[1ex]
\hline
\end{tabular}
\label{table:4jets}
\end{table}
\begin{widetext}
\begin{center}
\begin{figure}[h]
\epsfig{file=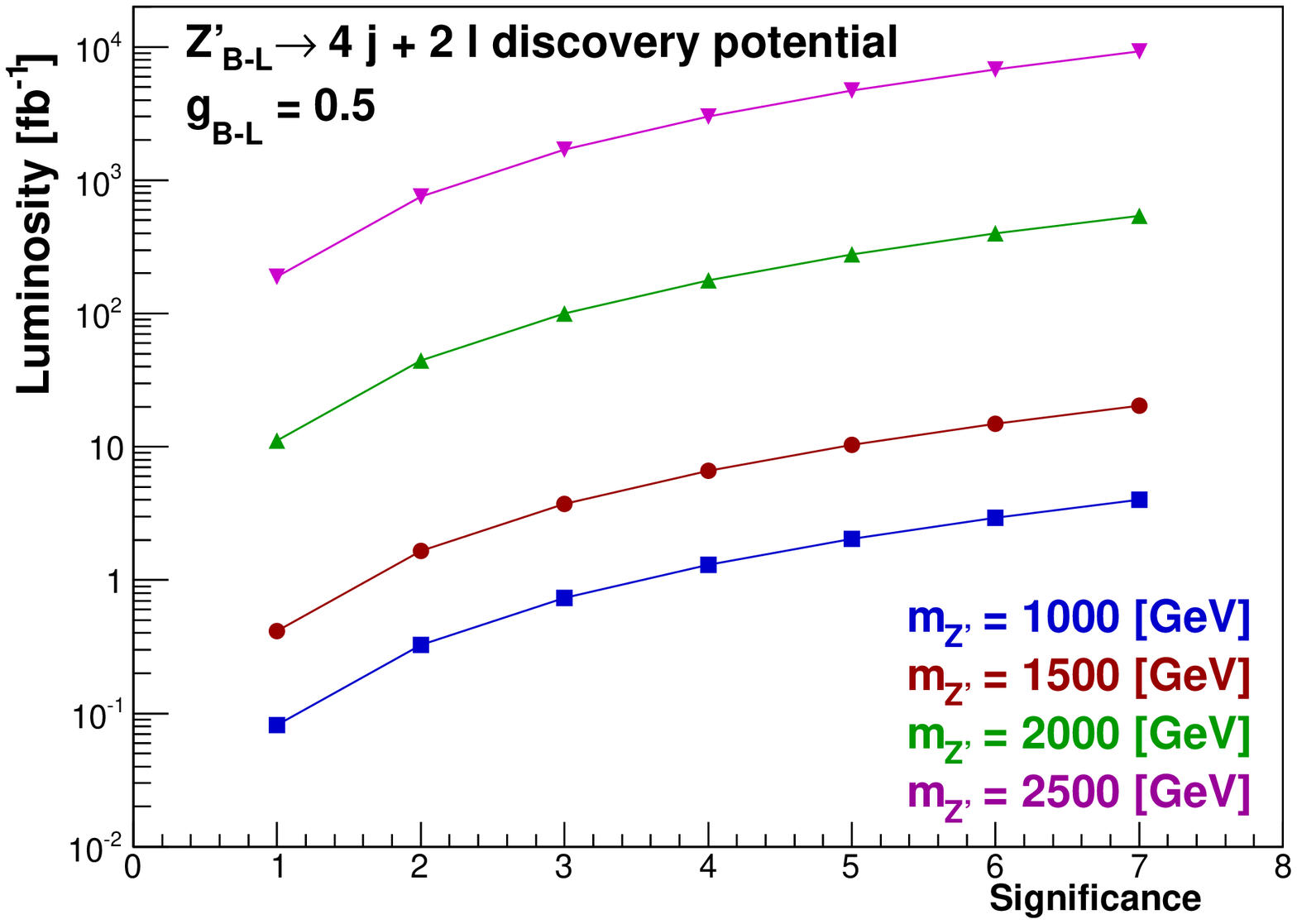, height=5.0cm,width=8.0cm,angle=0}~~~ \epsfig{file=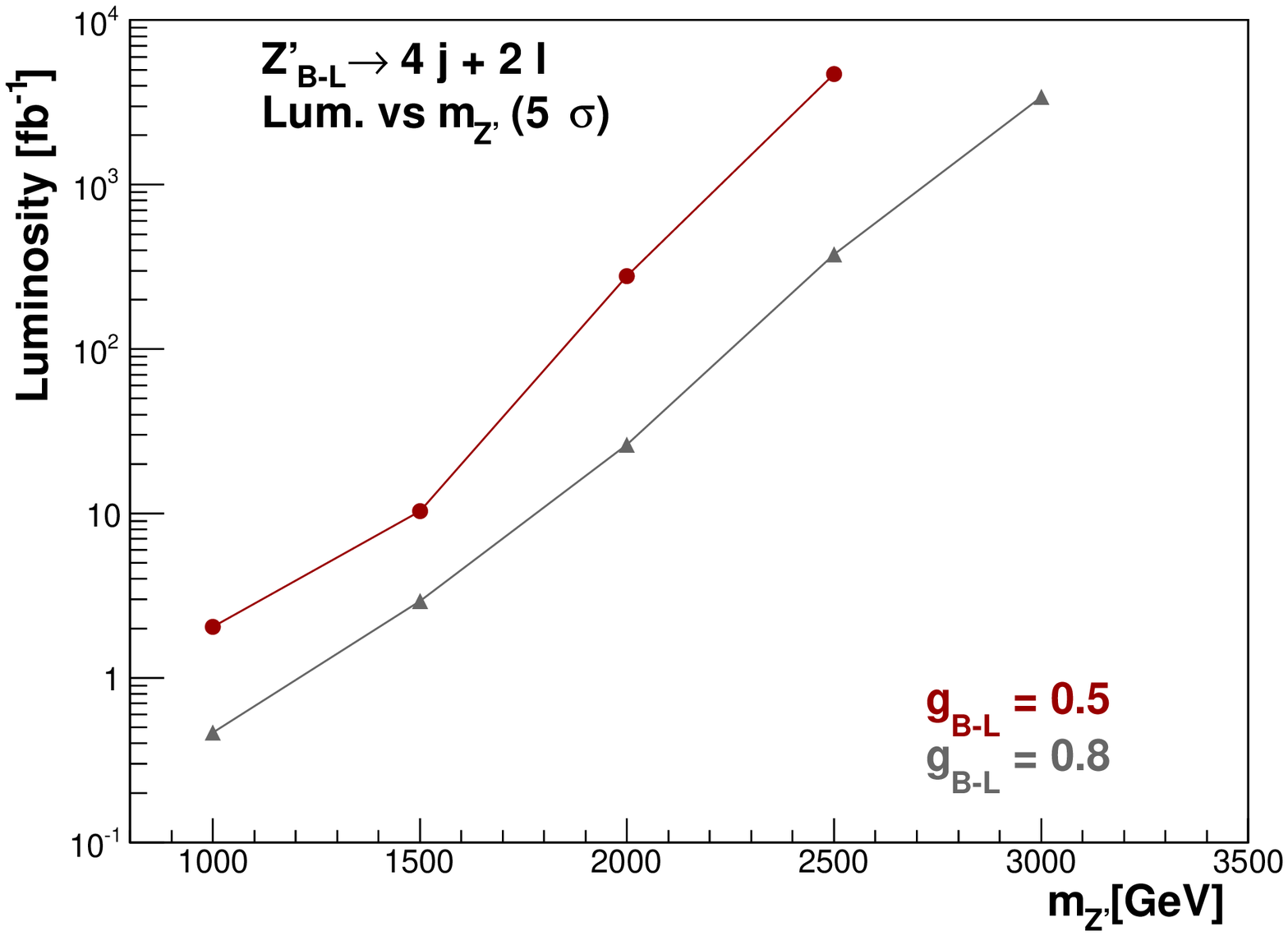,height=5.0cm,width=8.0cm,angle=0}
\caption {For the {\it4 j + 2 l} final state, the Integrated Luminosity of the data need (Left) for 1-7 $\sigma$ Statistical Significance discovery at  $g_{B-L}=0.5$ for different $M_{Z'}$, (Right) for a 5 $\sigma$ discovery as a function of $Mass_{Z'}$ at $g_{B-L} = 0.188$, $g_{B-L} = 0.5$ and $g_{B-L} = 0.8$. }
 \label{lumiForDisc4j}
\end{figure}
\end{center}
\end{widetext}

{\it iii) 3 l + 2 j + $\nu_l$ Final State}: in the semi-leptonic case where one of $W$'s decays hadronically and the other decays to $l+\nu_l$. The main background is $WZ+jj$. In the case of $W Z+jj$ associated production, three leptons can be generated from the subsequent leptonic decays of the two gauge bosons. In Fig.\ref{fig6} we show the invariant mass of $3l+2j+ \nu_l$ (left) for both of signal and background. Also the invariant mass of $2l+ \nu_l$ (middle) and that of $2j+ \nu_l$ (right). Again an additional $p_T >  150 GeV$ cut was set on the two $p_T$-leading leptons. Table \ref{table:semiLep} lists the number of event left after cuts for signal and backgrounds. 
In Fig. \ref{lumiForDisc3l2j} (lift panel) we plot the Integrated Luminosity of the data need for 1-7 $\sigma$ Statistical Significance discovery at  $g_{B-L}=0.5$ and $M_{Z'}$ = 1000, 1500 and 2000 GeV. In the right panel, we show the Integrated Luminosity of the data need for a 5 $\sigma$ discovery as a function of $M_{Z'}$ at $g_{B-L} = 0.5$ and $g_{B-L} = 0.8$.

\begin{widetext}
\begin{center} 
\begin{figure}[h]
\epsfig{file=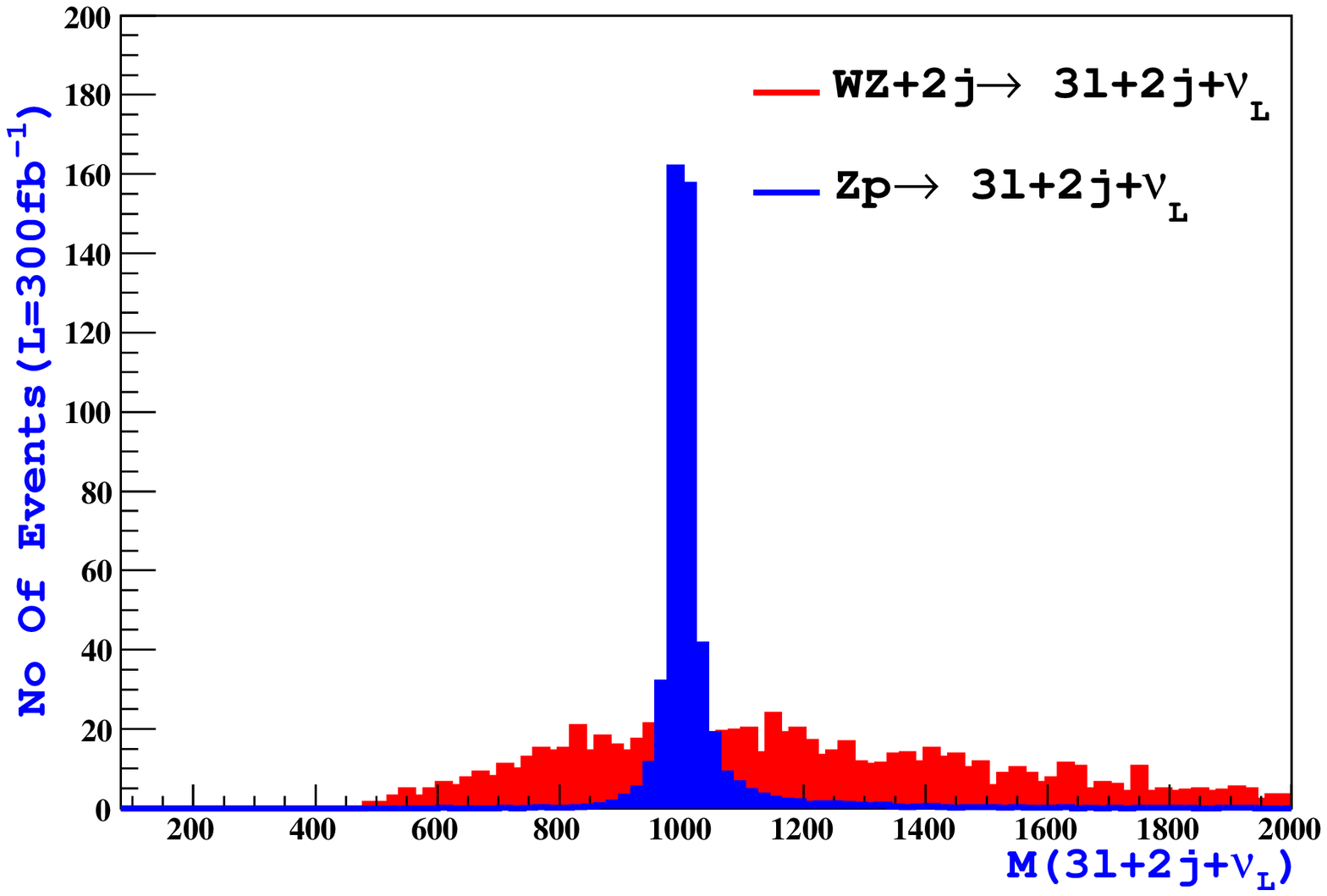,height=6.cm,width=7.cm,angle=0} \\
\epsfig{file=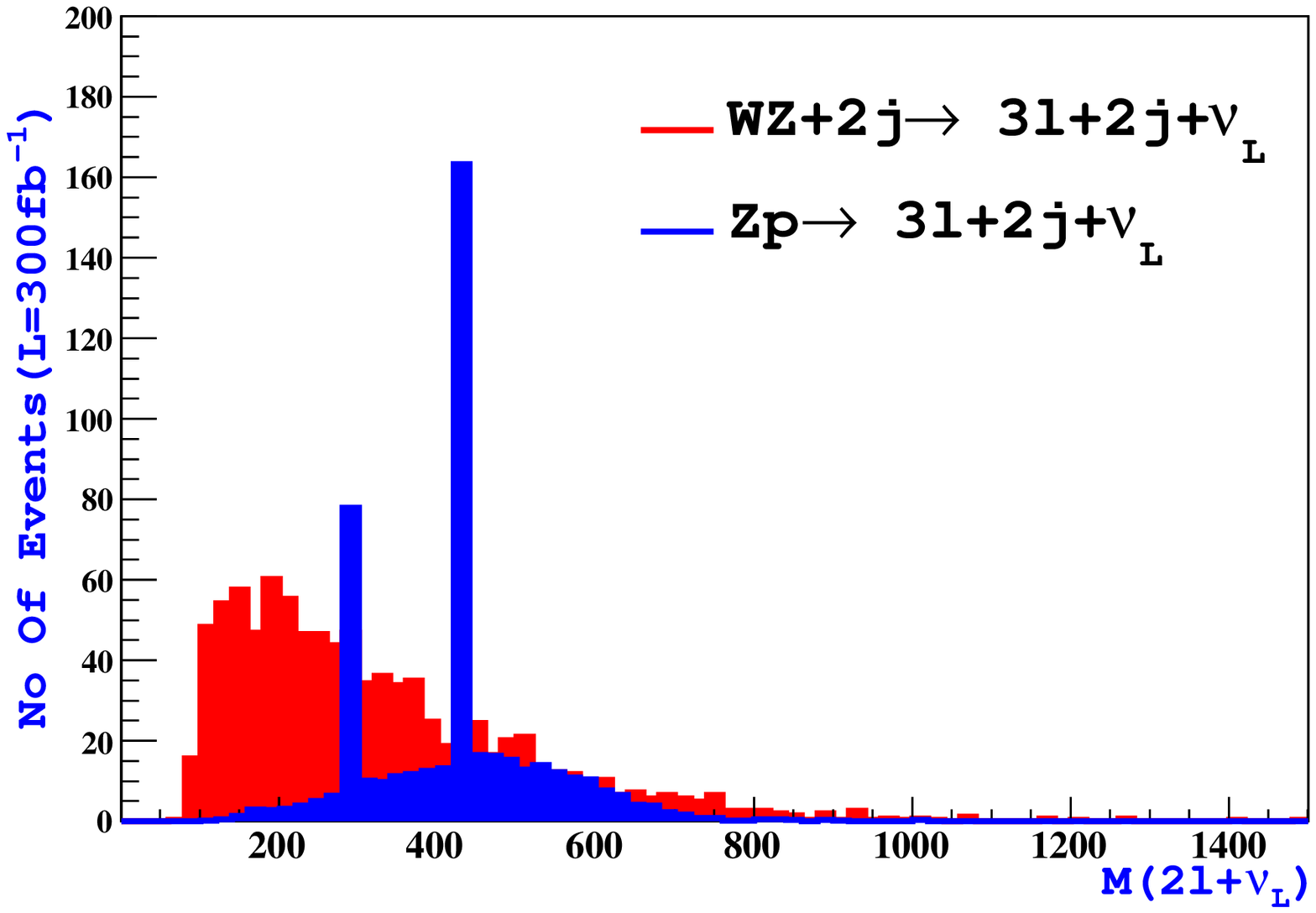,height=6.cm,width=7.cm,angle=0}
~\epsfig{file=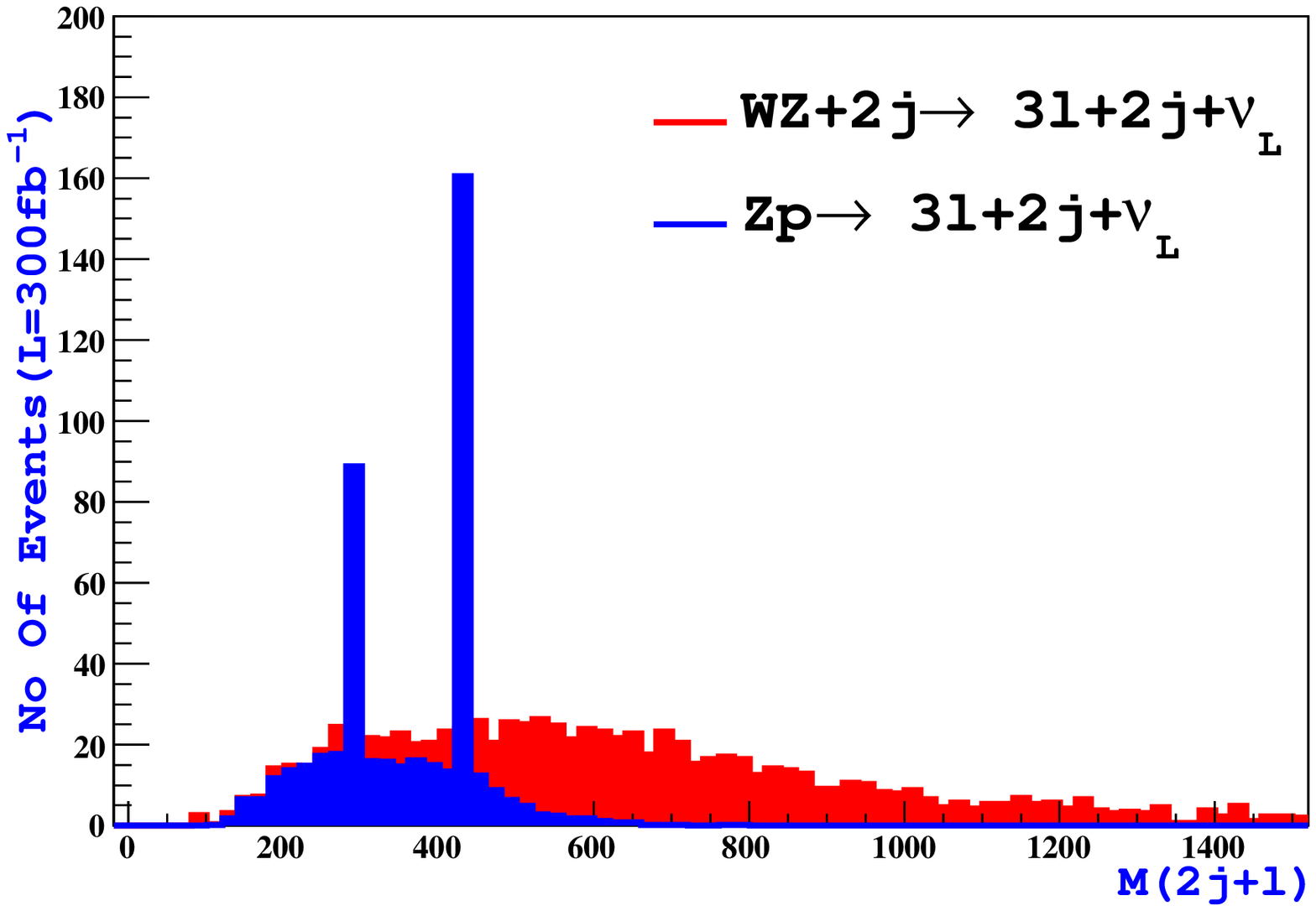,height=6.cm,width=7.cm,angle=0}
\caption{(Top) the invariant mass of $3l+2j+\nu_l$ distribution for both signal and expected background. (Bottom)
The invariant mass of heavy neutrino which decay to $2l+\nu$ (bottom-left) or to $2j+l$ (bottom-right) and its background}
\label{fig6}
\end{figure}%
\end{center}
\end{widetext}

\begin{table}[ht]
\caption{Number of events after initial set of cuts, in case $Z^\prime \to 3l+2j+\nu_l$ }
\centering
\begin{tabular}{c c c }
\hline\hline 
Cut\  & Signal\  & WZjj\   \\
\hline
Initial No of events \footnote{Weighted events, the initial generated number of events is 10 K events in siganl and 100 K events for background sample} &
769   & 37975  \\
$p_T > 150 GeV$ & 475 & 910  \\[1ex]
\hline
\end{tabular}
\label{table:semiLep}
\end{table}

\begin{widetext}
\begin{center}
\begin{figure}[h]
\epsfig{file=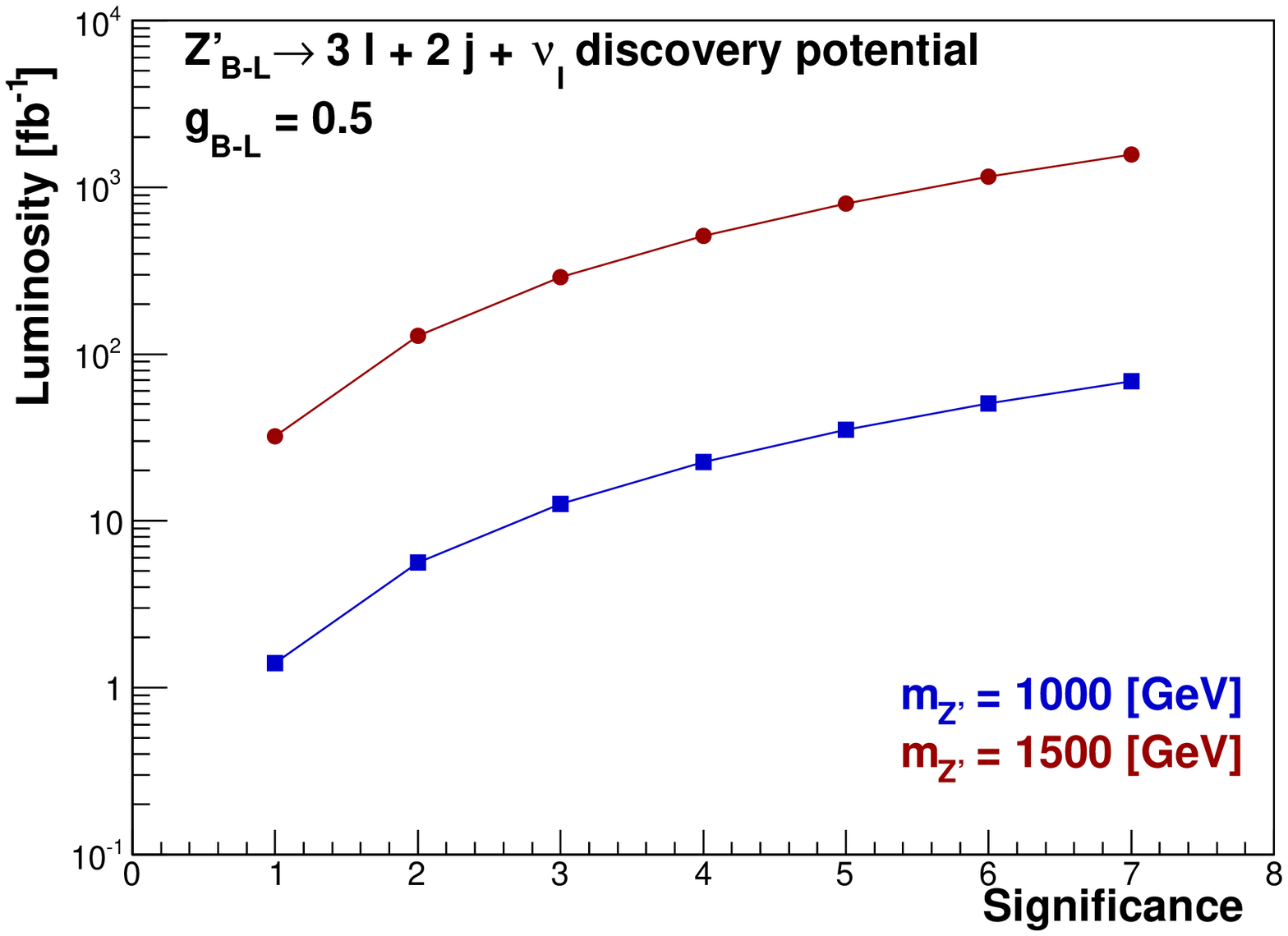, height=5.0cm,width=8.0cm,angle=0}~~~ \epsfig{file=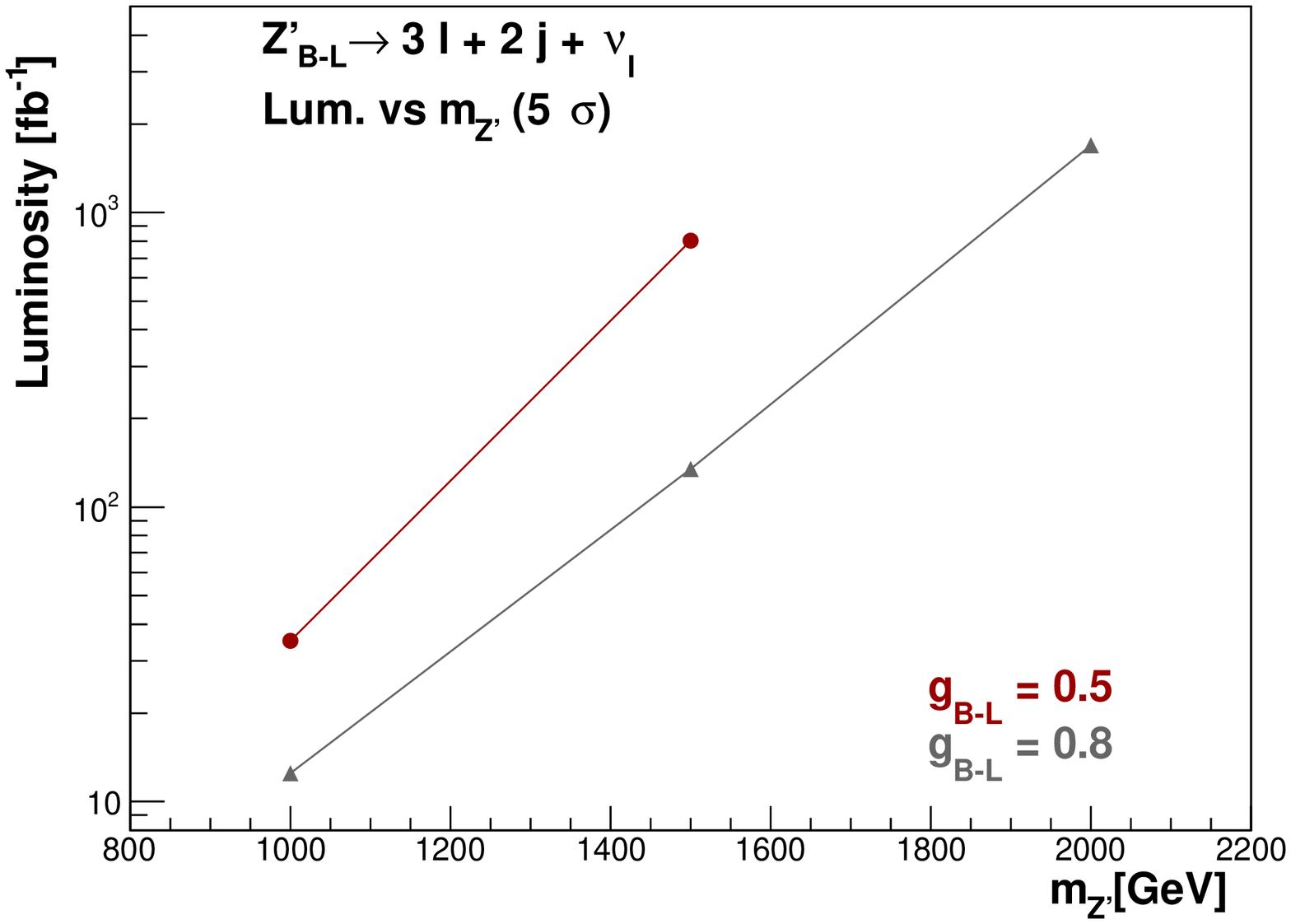,height=5.0cm,width=8.0cm,angle=0}
\caption {For the {\it3 l + 2 j + $\nu_l$} final state, the Integrated Luminosity of the data need (Left) for 1-7 $\sigma$ Statistical Significance discovery at  $g_{B-L}=0.5$ for different $M_{Z'}$, (Right) for a 5 $\sigma$ discovery as a function of $Mass_{Z'}$ at $g_{B-L} = 0.188$, $g_{B-L} = 0.5$ and $g_{B-L} = 0.8$. }
 \label{lumiForDisc3l2j}
\end{figure}
\end{center}
\end{widetext}

In summary, we have analysed the striking signatures of probing the heavy neuterinos, $\nu_h$,  and neutral gauge boson, $Z^\prime$, in TeV scale  $B -L$ extension of the standard model with inverse seesaw mechanism at the LHC. We have emphasised that in this type of models, where $Z^\prime$ may decay into new channels of heavy and light-inert neutrinos, the current experimental limits on $Z^\prime$ mass from LHC are relaxed. For instance, the limit on $Z^\prime$ mass can be lowered to $0.247$ of the current experimental value if $g_{B-L} = 0.5$. We provided detailed analysis for the pair production of heavy jeutrinos and their possible decay to four leptons and missing energy, due to two light neutrinos or to four jets and two leptons or to three leptons and two jets and missing energy due to one light-neutrino. In our analysis we have considered the following bench mark: $M_{Z^\prime}=1$ TeV, $M_{\nu_4}=M_{\nu_5}=287$ GeV, $M_{\nu_6}=M_{\nu_7}=435$ GeV, $M_{\nu_8}=M_{\nu_9}=652$ GeV. In addition the following cuts have been used: transverse-momentum $p_T \gsim 20 (10)$ GeV on final state jets (leptons), pesudo-rapidity $\eta \gsim 4(2)$ on jets (leptons), and separation between two jets (leptons) $R_{ij}$ ($R_{ll}) \sim 0.4 (0.2)$. In $4 j + 2 l$ channel and $3 l + 2 j + \nu_l$ channel an additional $p_T$ cut of 150 GeV on the two $p_T$-leading leptons is also applied. We showed that the $4l+2 \nu$ channel is almost free from the SM background, therefore it is the most promising decay channel for probing both $Z^\prime$ and heavy neuterinos at the LHC. 
We showed also the $Z'$ discovery potential at 14 TeV center of mass energy for the 3 decay channels under study.\\

\section*{Acknowledgments}
This work was partially supported by ICTP grant AC-80. The work of A. A. Abdelalim was supported by EENP2 FP7-PEOPLE-2012-IRSES grant. We would like also to acknowledge Florian Staub for the useful discussion. 


\end{document}